\documentclass[12pt,preprint]{aastex}

\slugcomment{}
\shorttitle{Mass Loss in Elliptical Galaxies}
\shortauthors{Athey et al.}

\begin{document}

\title{Mid-IR Observation of Mass Loss in Elliptical Galaxies}

\author{Alex Athey}
\affil{University of Michigan }
\affil{Department of Astronomy}
\affil{500 Church St.}
\affil{Ann Arbor, MI 48109-1090}
\email{alex@astro.lsa.umich.edu}

\author{Joel Bregman}
\affil{University of Michigan}
\email{jbregman@astro.lsa.umich.edu}

\author{Jesse Bregman}
\affil{Astrophysics Branch}
\affil{NASA Ames Research Center}
\affil{MS 245-6}
\affil{Moffett Field, CA 94035}

\author{Pasquale Temi}
\affil{NASA Ames}

\author{Marc Sauvage}
\affil{CEA/DSM/DAPNIA/Service d'Astrophysique }
\affil{C.E. Saclay }
\affil{91191 Gif-sur-Yvette Cedex }
\affil{France}

\begin{abstract}
Early-type galaxies exhibit thermal and molecular resonance  emission from dust that is shed
and heated through stellar mass loss as a subset of the population moves through the
AGB phase of evolution. Because this emission can give direct insight into stellar
evolution in addition to galactic stellar mass loss and ISM injection rates,
we conducted a program to search for this signature emission with CAM on ISO.
We obtained 6-15\( \mu  \)m imaging observations 
in six narrow bands for nine elliptical galaxies; every
galaxy is detected in every band. For wavelengths shorter than 9\( \mu  \)m,
the spectra are well matched by a blackbody, originating from the
K and M stars that dominate the integrated light of elliptical galaxies. However,
at wavelengths between 9\( \mu  \)m and 15\( \mu  \)m, the galaxies display
excess emission relative to the stellar photospheric radiation. Additional data 
taken with the fine resolution circular variable filter 
on one source clearly shows broad emission from 
9\( \mu  \)m to 15\( \mu  \)m, peaking 
around 10\( \mu  \)m. This result is consistent with the known, broad silicate
feature at 9.7\( \mu  \)m, originating in the circumstellar envelopes of AGB
stars. This emission is compared with studies of Galactic and LMC AGB stars to
derive cumulative mass loss rates. In general, these mass loss rates agree with
the expected \( \sim  \)0.8 M\( _{\sun }yr^{-1} \) value predicted by stellar
evolutionary models. Both the photospheric and circumstellar envelope emission
follow a de Vaucouleurs' R\( ^{1/4} \) law, supporting the
conclusion that the mid-infrared excess emission originates in the stellar component
of the galaxies and acts as a tracer of AGB mass loss and mass injection into
the ISM.
\end{abstract}

\keywords{Galaxies: elliptical; stellar mass loss; AGB stars; infrared emission.}

\section{Introduction}

Elliptical galaxies are comprised of old stellar populations in which low mass
stars evolve off the main sequence, eventually becoming white dwarfs. These
post-main sequence stars experience an asymptotic giant branch (AGB) phase where
up to 0.3 M\( _{\sun } \) is shed into the Interstellar Medium (ISM). As this
mass is ejected from the star, dust condenses out of the gas in a cool stellar
wind and forms an envelope at a few to ten stellar radii. (For a review of AGB
stars and circumstellar envelopes see \citet{AGB Star Review}). 
Dust acts as an agent
to this mass loss as it transfers radiation momentum from the star to the gas
through collisions. 
The stellar radiation absorbed by the dust heats it to temperatures from 300-1000K,
depending on the distance from the star.
Subsequently, the dust
cools thermally and has been detected and explored through the Infrared Astronomical Satellite 
(IRAS)
 and the Infrared Space Observatory (ISO) wide band filters
at 12, 25, 60 and 100\( \mu  \)m. 
(See \citet{Jura} for an early study of elliptical
galaxies with IRAS observations. 
See \citet{MidIR View of Galaxies} for a recent
review of mid and far IR emission from all types of galaxies.) 
In addition to
this thermal emission, broad line absorption or emission is observed in most
AGB stars at 10-12\( \mu  \)m and for oxygen-rich AGB stars, again at 20\( \mu  \)m 
(See \citet{Speck} for a recent
study of AGB mid-IR features). 
For older, low mass
stars found in elliptical galaxies, the AGB star envelopes have a high oxygen
content as opposed to the high carbon content present in high mass AGB envelopes, 
which occurs through inner layer carbon dredge up and expulsion. 
The theoretical models of dust condensation show
amorphous silicates (Si-O) to be the major constituent of the dust in these
oxygen-rich environments, where the stretching and bending of the Si-O bond
is responsible for the 10 and 20\( \mu  \)m absorption/emission. 
(Hereafter AGB, refers to the low
mass, oxygen-rich subclass of AGB stars.)
One of the surprises revealed by ISO are that in addition to amorphous
silicates, crystalline silicates are observed through many narrow band features
from 15-45\( \mu  \)m \citep{Ref A}. These observations have sparked renewed
interest in dust particle formation in AGB environments \citep{Ref B,Ref C}.
The dust composition, size distribution and other properties (albedo, dipole
strength, etc) define the interaction with the gas and ultimately control 
the observational consequences. Consequently, much laboratory work is being devoted to ``growing''
silicate dust particles and attempting to extract similar particles
from interplanetary dust particles brought to Earth through comets and meteorites
\citep{Ref D,Ref E}. 
These theoretical and lab studies have led to
a working model of AGB star dust formation and emission.

Although much work has been done on the study of individual AGB stars, population
studies of these AGB dust features in early-type galaxies is relatively unexplored
due to the faint nature of the emission involved. In an elliptical galaxy, it
is the sum of many individual oxygen-rich AGB stars that will produce in aggregate
a similar feature to individual AGB stars. This feature can be used to
confirm this general stellar evolution picture and determine the mass loss rate
into the entire galaxy. An effort to detect this mid-IR excess
in early-type galaxies was first carried out by \citet{KGWW} (KGW
hereafter) who used the IRAS All Sky Survey coupled with ground based 10\( \mu  \)m
data to search for the dusty component of the ISM in nearby elliptical galaxies.
Because this emission is quite faint, these galaxies were detected at 
low signal-to-noise ratio. 
Nevertheless, they observed excess emission at 12\( \mu  \)m
relative to a derived stellar continuum, indicating emission from
the circumstellar dust envelope. KGW scaled the emission from Galactic
AGB stars to their signal and they determined galaxy-wide mass loss rates
of $\sim$0.7M\( _{\sun }yr^{-1} \). Also, they show that the dust emission
is extended on the scale of the galaxy, ruling out the hypothesis that the dust
emission originates only in the nuclear regions. However, many of the galaxy
detections were at the 2\( \sigma  \) level and the coarse IRAS sampling (1\( \arcmin  \)x5\( \arcmin  \))
prevented good spatial sampling of the galaxies.
Also in the KGW study, the sample
of Galactic stars used to calibrate the observed excess did not have well determined
distances, leading to errors in the scaling relation. Subsequent to the KGW
work, significant advances in understanding have been made in the study of individual
AGB stars with detailed observational monitoring programs in addition to improvements
in the modeling of the mass loss mechanism and the subsequent radiation (\citet{Whitelock},
hereafter W94; \citet{Trams  Obs} and \citet{Jacco Mdot}, hereafter L99). These studies are essential because it
is necessary to characterize the emission from individual AGB stars in order
to understand the collective AGB emission from integrated galaxy light. 

There are a number of key issues that IRAS was not able to
address effectively due to instrumental limitations. First, it is important
to definitively detect and characterize this MIR excess. Specifically, is there
a spectral excess that peaks around 9.7\( \mu  \)m and is consistent with a
physical source of an AGB star plus circumstellar dust envelope? Second, it
is useful to determine if the excess emission is similar from galaxy to galaxy,
as would be expected for similar coeval populations. If there are significant
differences, this emission could be used as a diagnostic tool for star formation or
history, metallicity and age of these populations. Finally, it is important
to determine if this excess follows a de Vaucouleurs law, as would be expected
if the circumstellar envelopes around dying stars are the only source of the
MIR excess \citep{de Vaucouleurs}. We have used the CAM instrument on ISO to
address these issues because it offers: 1) excellent sensitivity in six narrow-band
filters that cover the 6-18\( \mu  \)m region 2) high spectral resolution imaging
using the CVF in the same spectral region and 3) spatial resolution of 6\( \arcsec  \)
over a field of view of 3.2\( \arcmin  \). Here we report upon and analyze
ISOCAM observations for nine nearby early-type galaxies.

In Section \ref{Observations} we describe the observations made by ISO with the CAM
narrow band filters and CVF.  We report on the extensive measures we undertook to 
process and understand the CAM detector data in Section \ref{data}.  The mid-IR emission
detected in these nine galaxies is described and characterized in Section
\ref{results}.  In Sections \ref{E12} through \ref{comparemdot} the mid-IR excess
emission is compared to that of Galactic and LMC AGB stars to derive a scaling relation, revealing cumulative mass
loss rates for the observed galaxies.  This observed mass loss rate is reasonably matched to
theoretical predictions as described in Section \ref{theory}.  Section \ref{Light Profile Section} examines
the radiation vs radius profile of these galaxies in different wavelength regions.  
We summarize and add final thoughts in Section \ref{conclusion}.

\section{\label{Observations}Observations}

We observed nearby elliptical galaxies with a range of optical and environmental
properties in the mid-Infrared (MIR) with the Infrared Space Observatory (ISO)
\citep{ISO inst paper}. Nine galaxies
were observed in the ISO accessable regions of the sky, resulting in a statistically 
incomplete, but unbiased sample selected from a complete survey of 
optically bright early-type galaxies observed in the X-rays \citep{Brown & Bregman}.
Table \ref{Galaxy Properties Table} gives basic data about the galaxies in the sample.
The CAM imaging instrument operated with a 
3.2\( \arcmin  \) field of view and a scale of 6\( \arcsec  \) 
per pixel \citep{ISOCAM Inst Paper}.
Observations were made in long wavelength (lw) filters 4-9, corresponding to a 6-15\( \mu  \)m
wavelength coverage. Table \ref{Exposure Time Table} reports central wavelength
and bandwidth information for these ISOCAM filters \citep{ISOCAM Data Handbook}.
As is typical with ISO, detector flux stabilization observations were taken prior to each
target observation. Galaxies and their exposure times, including stabilization
exposure times, are also listed in Table \ref{Exposure Time Table}. Each filter
exposure is composed of four pointings with a 1/2 pixel (3\( \arcsec  \))
raster movement between them. In every band all the galaxies are detected with
a S/N of at least 10.

In addition, a Circular Variable Filter (CVF) observation was obtained of NGC 1404.
This mode of observation covered the wavelength range from 6.22\( \mu  \)m
to 14.2\( \mu  \)m at a resolution of \( \lambda /\triangle \lambda  \) \( \sim  \)40
\citep{ISOCAM Data Handbook}. We scanned the spectral range in 38 steps and
both the upward and downward scan directions. Each of the 10 frames per step
position was taken with a 3.2\( \arcmin  \) FOV and an integration time of
10 seconds.

\section{\label{data}Data Reduction}

ISO uses a Si:Ga detector which has the typical electronic detector signatures
of bias level, pixel-to-pixel variation, and large scale sensitivity variation.
In addition, these devices respond to any change in illumination with a transient
period before reaching a stabilized electronic signal. The transient behavior
of these detectors can be difficult to consistently characterize because the
transient period is a function of the incident photon flux. Also, many of the
astronomical observations are completed before the stabilization period has
been achieved. Furthermore, the detector's 32x32 array of ``pixels'' are
not physically distinct. Instead, electric fields are laid across the substrate
to create regions that behave like pixels. Cross-talk between the pixels can
be seen in the data as some pixels steal the electric charge from neighbors.
Because of these peculiarities, the data reduction process requires two non-standard
corrections. To correct the transient effect, models employing solid-state
physics theory are matched to the detector response. This correction process is imperfect
and sometimes results in a gross over-correction of the data, usually seen as
a singular spike in time. Most of the cross-talk and these singular spikes are
removed manually, inspecting each frame in the data cube (x, y, time). The remaining
anomalies contribute little to the corrected data as they are smoothed through
time averaging. We carefully studied all the corrections applied to our data
in order to most accurately reconstruct the incident radiation and to have a
complete understanding of the uncertainties within that product.

\subsection{ISOCAM - LW4-LW9}

The ISO data were reduced using CAM Interactive Analysis (CIA) version 3.0 (October
1998) developed by the instrument team.\footnote{%
The ISOCAM data presented in this paper were analysed using CIA, a joint development
by the ESA Astrophysics Division and the ISOCAM Consortium led by the ISOCAM
PI, C. Cesarsky, Direction des Sciences de la Matiere, C.E.A., France. 
} First, the data were grouped into individual filters. In these filter groupings,
we retained the pre-observation grade pointings that occur prior to program
observations for flux stabilization, suppling as much information as possible to the transient correction
(below). We subtracted a dark constructed from ISO calibration data and adjusted
for the dates of our observations. Outlined below are the extensive measures that were taken to minimize the 
transient effect in our data. 

There are two important signatures that need to
be removed from the data; an unwanted physical source (cosmic rays) and a detector
electronic effect (variable gain or transient effect). These two signatures
are not independent as every cosmic ray event
decays exponentially back down to the background level. In order to completely
remove the impact that cosmic rays had on the data, it was necessary to identify
and mask both the cosmic ray event and its subsequent transient trail. For our
automated deglitching routines, we split the data array into two regions: a
central or source region centered on the galaxy with a five pixel radius
and an outer background region. In the background region, an aggressive
(low sigma rejection limits) software deglitching routine based on both temporal
and spatial dimensions removes most of the non-stabilized signal. 
In the source region, a temporal deglitching
method with large sigma rejection limits retains the signal when the satellite
moves between raster positions. Next, additional
cosmic rays or glitches overlooked by the automated routines were masked. 
The entire timeline for each pixel in the data cube was reconstructed, including pre-observation
grade data used to start stabilizing the detector. 

After removing all the cosmic rays and their transient trails from the data,
we corrected for the transient effect. Of all the transient
correction methods available, the Fouks-Shubert (F-S) method
best corrected our transient data to the stabilized flux levels. This method
is based on the Fouks-Shubert theory of Si:Ga detectors and is used to create
a non-linear differential equation which describes the physics of the p-type
photo-conductor in response to incident radiation. This theory models the non-linear
response and decay of signal as it is received by this type of detector. The
solution to the equation can be inverted and applied to the data, thereby yielding
the stabilized fluxes \citep{trans corr}. The mathematics used to solve the
F-S equation are non-trivial and many solving algorithms have been developed
to tackle the problem. The algorithm that worked best for our data was external
to CIA3.0 and developed by Herve Wozniak. This algorithm was subsequently incorporated
into CIA4.0

\begin{figure}
\plotone{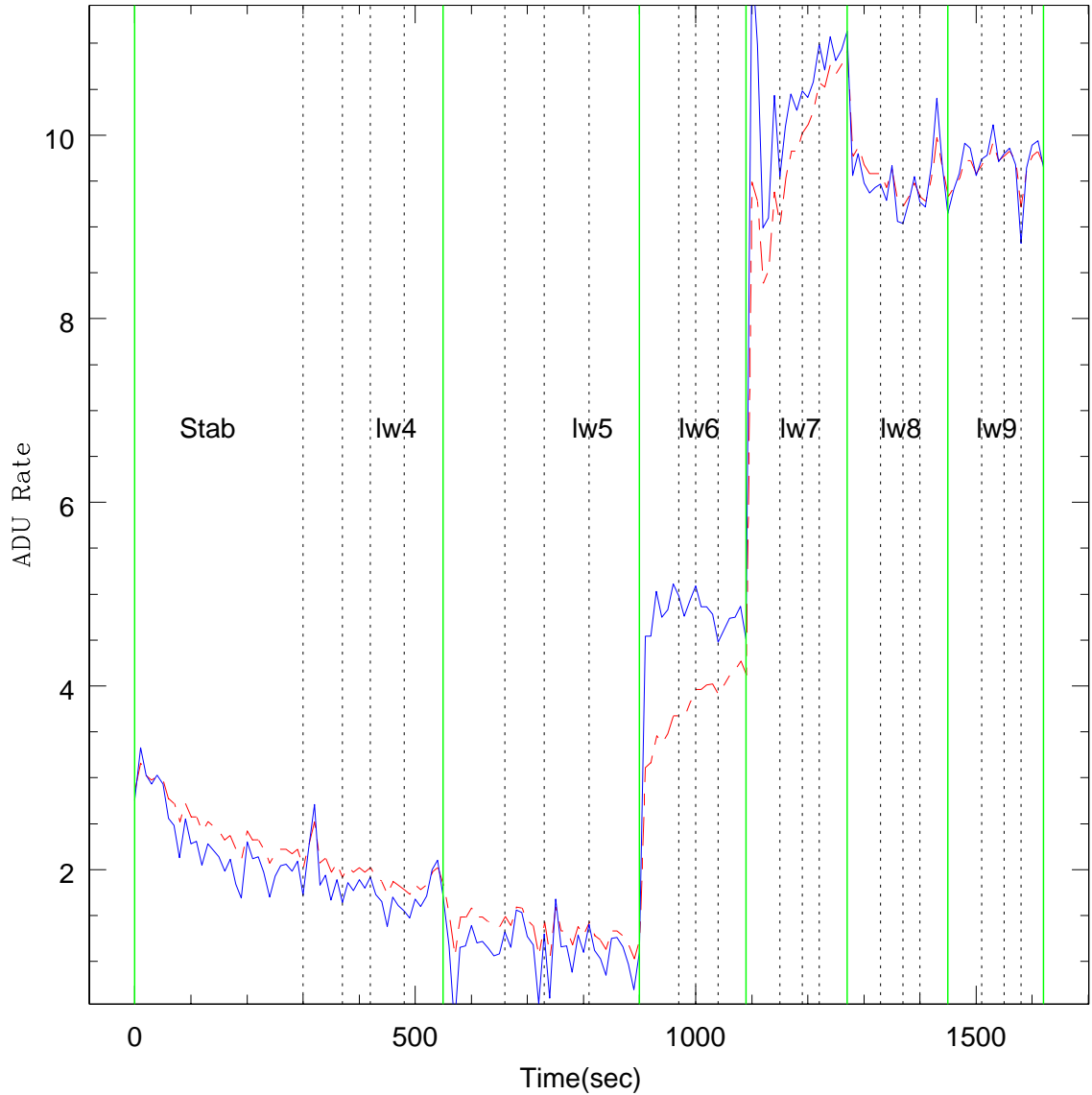} 
\caption{\label{sky-trans}
The electronic detection history of a typical sky pixel.
Each filter is divided by a dotted
line and individual pointings within a filter are separated by dashed lines.
The lw4 filter has an additional pointing used for flux stabilization of the
detector, labeled ``Stab''. The dashed line follows the non-transient corrected data, while the
thin solid line traces the F-S corrected data. In this wavelength regime the
zodiacal light is the dominant background source and as a result large changes in the background
level occur as the filters are changed. This is particularly visible in the
change from filter lw5 through lw6 and into lw7. In the lw6 band the transient correction 
seems to do quite well; all of the
corrected data are at a similar level (\( \sim  \)4.7\( \pm  \)0.2). The un-corrected
data have a lower average (\( \sim  \)4.0) and a clear positive derivative.
}
\end{figure}

On a pixel by pixel basis, it was difficult to judge the effectiveness of any
given transient correction. However, 
large regions, such as source pixels and sky pixels, display the trends of the
data and the transient correction with clarity. The effectiveness
of a transient correction method was judged by examining a pointing that had been allowed
sufficient time to approach a stable value by the third and fourth rasters,
but was transient in the first two. We compared the corrected values of the
first two pointings to the last two stabilized levels. 
Finally, the transient correction is imperfect and frequently leaves spikes
in the data (e.g. lw7 in Figure \ref{sky-trans}). We meticulously searched
the data and manually removed these singularities in the solution.

With the data corrected for transient behavior, the data cubes were flattened and combined 
for individual filters. Because the small movement between rasters
1/2 pixel inhibited us from producing a flat from our own data, the
library flat from the calibration files of ISO was used. The pointings were corrected for
distortions and the four rasters were combined with calibration tasks in CIA. The
flux units were converted to mJy with the standard conversion incorporated in
CIA3.0.

The transient correction was the most uncertain step in our reductions. In order
to gauge the level of this uncertainty, we split each filter observation into
its individual raster observations, giving us four independent observations
per filter. Aperture photometry was performed on each of these raster images
in each filter and the standard deviation about the mean determined the error in the data.
(See Table \ref{fluxes} for mean fluxes and errors.) In cases where the statistical
noise in the sky was larger than deviation about the mean, these larger errors
were used.

From these data, a six-band spectral energy distribution, SED, was extracted 
as well as surface brightness profiles. For the spectral energy distribution, 
the flux was summed within a 30 arcsecond radius centered on the galaxy and then subtracted
off from a background determined in the outer parts of the image. The vignetting
was evident on some image edges so we excluded the outer 3 pixels from all of
our background measurements. The surface brightness measurements were determined
in annuli centered on the galaxy out to the edge of the chip, with radial increments
of one pixel. The background for some of the galaxies was within an effective
radius and galaxy contamination was a concern. However, the SED of the background
region matched recent ISO observations of the zodiacal light, giving us confidence
in our subtracted fluxes \citep{Ref G}. Furthermore the profile measurements
will not be affected as no background is subtracted, and, in fact, little to
no increase in flux over \( r^{2} \) is seen in the profiles of the background
region.

\subsection{ISOCAM - CVF (Circular Variable Filter)}

The CVF data were processed independent of the CAM because the exposure times
are significantly shorter, filters much narrower and flux levels reduced
from the narrow band passes. These conditions force different methods to be used
to eliminate the instrument's electronic signature. Data reduction followed
the standard CVF steps: after dark current subtraction we cleaned the data cube
of glitches using both the CIA multiresolution median transform algorithm and
a manual deglitching method by inspecting the cube frame by frame. As with the
narrow band filter data, the Fouks-Schubert model for stabilization was used
and a jitter correction was applied to the data. 

We devoted effort to producing an appropriate and accurate flat field and background
subtraction procedure. Stray light from extended sources may affect the flat-field
quality of the reduced images. The standard flat-field procedure on CVF images
produces a deviation from flatness, around 10\% on average, which limits the
ability to image faint extended sources and to perform accurate photometry.
The characteristic double-lobe CVF flat field pattern, present in the data after
this flat field process, can be removed by subtracting an appropriate zodiacal
light cube and by using a more appropriate detector flat field. However, since
neither cube was available from the standard distribution of calibration files
at time of reductions, we produced our own. We constructed a zodiacal light
cube for both sectors of the lw CVF by interpolation, starting from 8 zodiacal
images taken with the CVF at fixed wavelengths within the wavelength range of
our measurements. We were able to remove the zodiacal light by scaling this
cube to the intensity measured off-source in our observations, and subtracting
the scaled cube from our measurements. Flat fields were produced in a similar
manner. From the data archive we recovered all the detector flats which were
taken with the 6\( \arcsec  \) FOV through narrow band filters. These flats
were interpolated to match the wavelengths in our data set, and each data frame
was divided by the appropriate flat. The final calibrated data for NGC 1404,
the only galaxy observed by CVF in our program, is reported as the background
subtracted flux within within  30\( \arcsec  \) radius in Table \ref{fluxes-cvf}.

\section{\label{results}Results}

\begin{figure}
\plotone{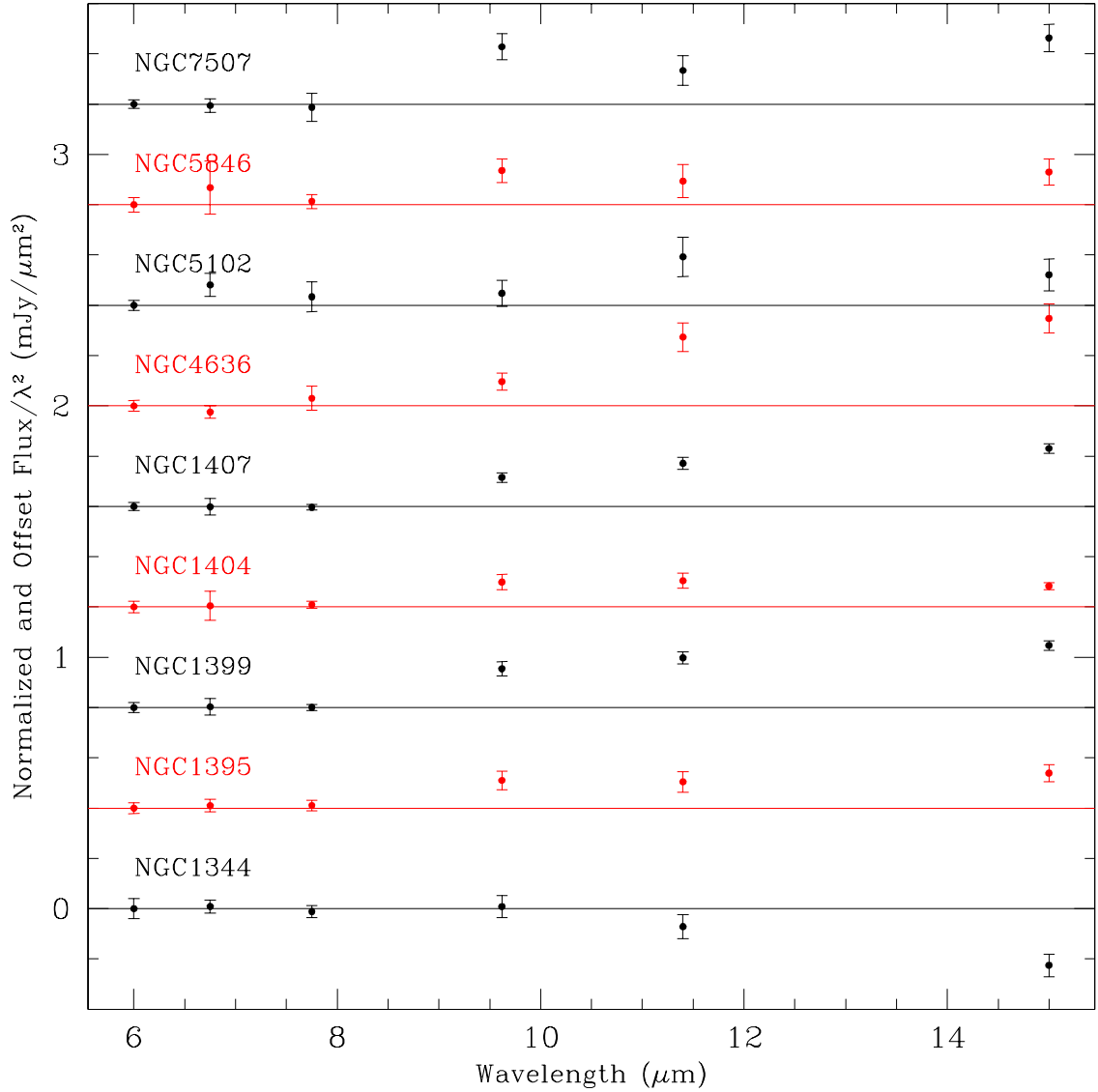} 
e\caption{\label{Galaxy SEDs}Galaxies NGC 1344, NGC 1395, NGC 1399, NGC 1404, NGC 1407, NGC 4636, NGC 5102,
NGC 5846 and NGC 7507 SED with blackbody fits.  The flux is normalized and divided by wavelength squared.  
At these wavelengths a blackbody of T$\sim$3,000K is a straight line in these units and fit to the first three data points.  Offsets are 
applied to distinguish individual galaxies.  Clear excess emission over a blackbody in the 11-15 \( \mu  \)m
can be seen in all galaxies except NGC 1344.}
\end{figure}

Elliptical galaxies are comprised of mostly old K and M-type stars. Because
the spectral energy distribution of these stars follows the Rayleigh-Jean's
part of a blackbody distribution from 6\( \mu  \)m to 7.75\( \mu  \)m, we
fit a cool (T$\sim$3500K) blackbody to our first three data points
(See Figure \ref{Galaxy SEDs}). These data points are also well fit by IRAS
Low Resolution Spectrograph (LRS) data of late type stars, confirming that our
choice of a blackbody function is appropriate for this region. When this blackbody
law is extended to longer wavelengths, a clear excess over stellar emission
can be seen in wavelengths 9.62\( \mu  \)m-15.0\( \mu  \)m for all of our
galaxies, except NGC 1344. The same results apply when a ratio of the two different
emission regions is taken and compared to the expected color from a blackbody
or LRS spectra of late type non-mass-losing stars. This excess emission coincides
with the 9.7\( \mu  \)m silicate dust band that originates in the circumstellar
envelopes of AGB stars. It has been established that the emission from this
dust band, when calibrated, can reveal the mass loss rates for AGB stars \citep{Ref F,KGWW,Whitelock}.

\begin{figure}
\plotone{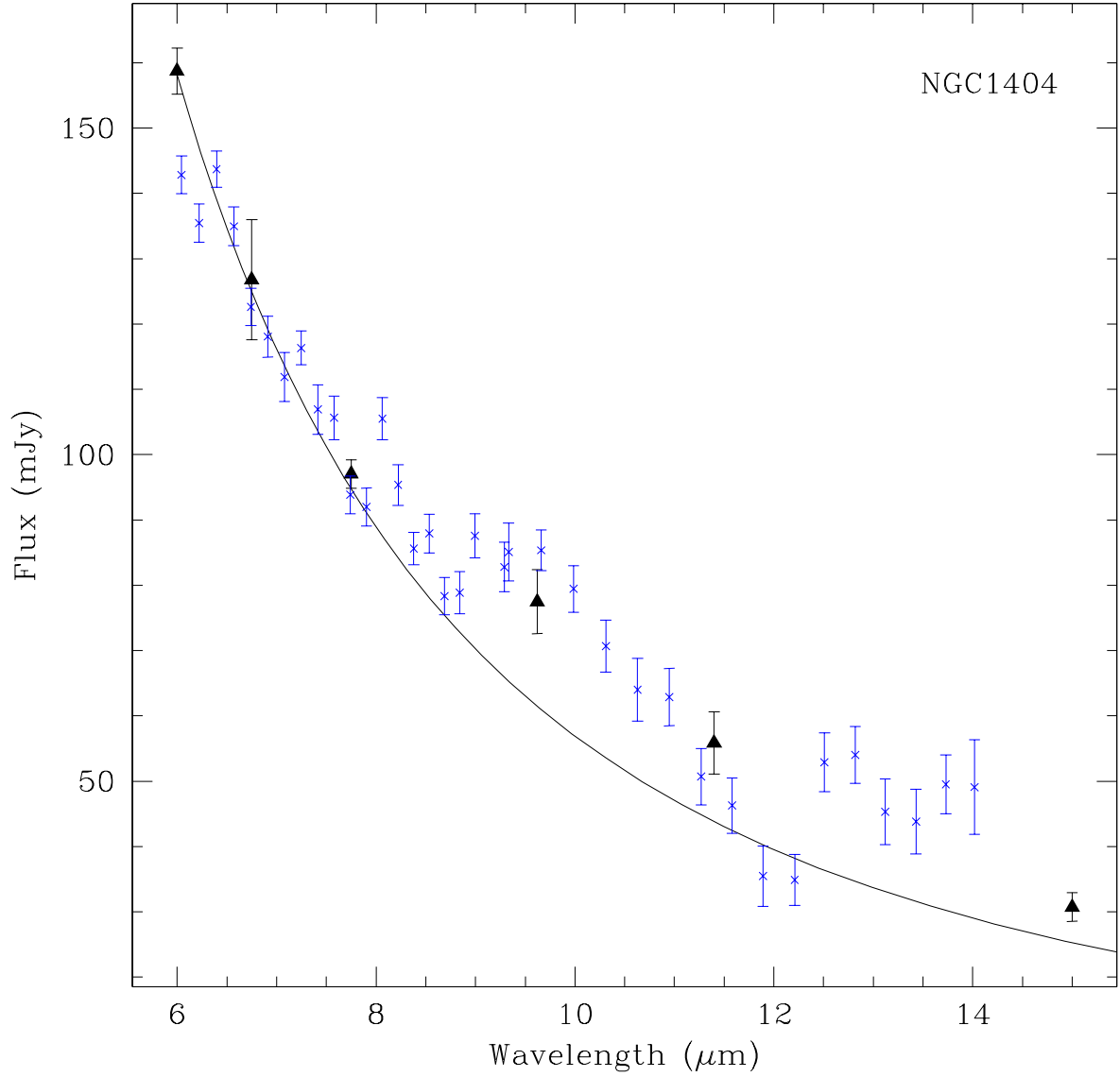}
\caption{\label{NGC1404 Spectra}Galaxy NGC 1404 spectral energy distribution from CAM and CVF data with a blackbody
fit.  Emission feature at 9.7\( \mu \)m is consistent with oxygen-rich AGB silicon dust feature seen in many nearby AGB stars.}
\end{figure}

In the one galaxy for which we have CVF data, NGC 1404, the results are consistent
with the narrow band SEDs (Figure \ref{NGC1404 Spectra}). A stellar
continuum in the shorter wavelengths, 6-9\( \mu  \)m is present, with an emission feature
from 9-11\( \mu  \)m. This feature is well matched by IRAS LRS data from late-type
mass losing stars, such as \( gHer \), and it is also coincident with laboratory
spectra of silicate grains \citep{Ref H,Ref I,Ref J}.

In the mid-infrared, it is also possible to detect poly-cyclic aromatic hydrocarbons
(PAHs) as they have characteristic emission in this wavelength region. However,
we see no evidence of PAH emission in the strongest band at 7.7\( \mu  \)m with a typical upper limit
of five mJy.

\subsection{\label{E12}12\protect\( \mu \protect \)m Excess}

In order to compare our data with the literature, it useful to define
a ``12\( \mu  \)m Excess,'' (\( E_{12} \)) which is the broad-band 12\( \mu  \)m
flux minus 38\% of the K-band(2.2\( \mu  \)m) flux, where the 12\( \mu  \)m
flux can be either the IRAS 12\( \mu  \)m band or the ISOCAM lw10 filter, whose
responses are nearly identical \citep{IRAS PSC,ISOCAM Data Handbook}.
Because this 12\( \mu  \)m band includes all of the flux expected from the
9.7\( \mu  \)m silicate feature, when the stellar contribution is subtracted,
the remaining flux should be circumstellar dust emission. 

Because we did not have broad-band ISOCAM lw10 measurements, we used the ISO
system response data to develop a flux conversion between the narrow band filters
lw6-lw9 to lw10. The relation we determined was 
\begin{equation}
lw10 = 0.06 lw6 + 0.22 lw7 + 0.55 lw8 + 0.17 lw9,
\end{equation}
where lw\# indicates the flux measured through each filter. We also calculated
a K(2.2\( \mu  \)m) flux assuming the lw4-lw6 flux was stellar in nature and
extrapolated in the Rayleigh-Jeans limit. Rather than use K magnitudes reported
in the literature, we found that this extrapolation to 2.2\( \mu  \)m was a
better comparative measurement for our galaxies because of the different apertures
and systems used in the literature. Also, the absolute calibration of ISO is
not as well-determined as the internal calibration, giving us further reason
to rely on internal calibrations rather than use an absolute system.
In Table \ref{K-[12]} we report the log of the ratio of the calculated lw10
filter to extrapolated K-band flux as a K-{[}12{]} color for these galaxies.
A hot blackbody (T$\sim$10,000K), which is equivalent to an A0 V star at these wavelengths,
sets the zero point of the K-{[}12{]} color. 
The errors listed are propagated from the random errors
in the data and quantities from the literature

\subsection{Mass loss Observations}

In order to determine mass loss rates for our galaxies, we scaled individual
mass-losing AGB stars to match our data. These individual AGB stars have semi-empirical
mass loss rates determined by matching models to observational data. The two
most common observations are CO line profile data and IR photometric measurements.
The IR photometry detects circumstellar shell fluxes, while optical-waveband
data are typically used to track the stellar component. The photometry data are
compared to spectral energy distribution models and the CO data are matched
to model profiles where a combination of velocity and column density reveals
the mass loss rate. From various literature searches, there is generally an
uncertainty of a factor of \( \sim  \)4-5 in these models and there are parameters,
such as metallicity, gas to dust ratio, and age, that play a large, but relatively
undetermined role. \citet{Ref K} compared a Galactic, an LMC, and an SMC AGB
star all with similar periods and concluded that AGB mass loss rates 
increase with metallicity. However, too little data were available to derive
an exact relationship. Theoretical studies of mass loss have also noted this
effect but have been unable to quantify the exact nature \citep{Ref L}. \citet{Ref M}
compares three decades of MIR spectra of AGB stars and finds evidence for significant
changes in the silicate feature of AGB stars within this short timescale.
Our approach to these uncertainties was to explore each of the mass loss scaling
relations available and present the range of determinations as likely uncertainties
in the current models and methods. For each of the methods we compared our integrated
light colors/relative fluxes to individual mass-losing stars.

\subsubsection{\label{MDOT KGWW}Previous Mass Loss Relations}

\begin{figure}
\plotone{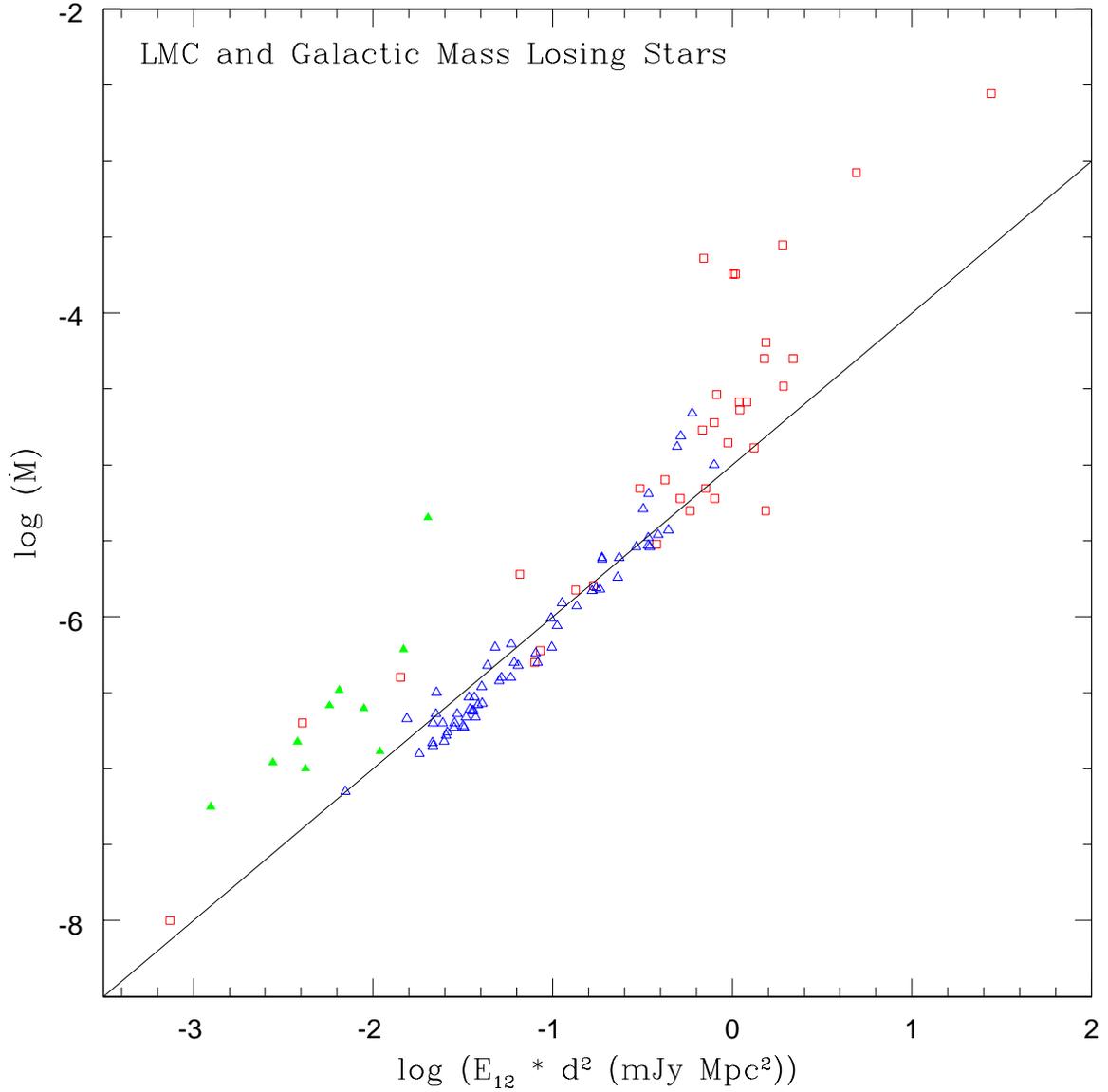}
\caption{\label{Stars Mdot}Mass loss rates for individual AGB stars as a function of 12\( \mu \)m
Excess (\protect\( E_{12}\protect \)), scaled to distance squared. Triangles
are Galactic stars, with filled triangles from KGW, corrected for Hipparcos
determined distances, and open triangles from W94. LMC stars from
L99 are represented by open squares. Solid line shows a linear 
relationship.}
\end{figure}

As a first attempt at calibrating our 12\( \mu  \)m excess, \( E_{12} \),
to a mass loss rate, we re-examined the Galactic AGB stars KGW used
to scale their emission. The main difficulty in AGB star studies is determining
their distances. KGW assumed an absolute magnitude of -8 for all of
these stars to obtain the distances. Subsequently, Hipparcos measured distances
for some of these stars, although some are at the limit of detectability and
thus have large associated errors. The KGW relation scaled to
the new Hipparcos distances and corrected for the Lutz-Kelker bias (Figure
\ref{Stars Mdot})\citep{lk correction,lk} 
allowed us to derive an updated mass loss rate relation of the form:  
\begin{equation}
log \dot{M} = log(E_{12}*d^2) - (4.44\pm 0.10).
\end{equation}
The Hipparcos
distances lead to a correction of +0.41 in the zero point of this relation from
the original. The average Hipparcos distance was 55\% smaller than the assumed
distances, leading to the large correction. We apply this calibration to our
12\( \mu  \)m Excess, as shown in Table \ref{K-[12]}, in the column labeled
``KGW''.

\subsubsection{\label{Mdot Whitelock}Galactic AGB Stars}

A systematic study of Galactic AGB stars has been carried out by W94,
who obtained multi-epoch, ground-based K-L observations which were compared
with IRAS observations. A key feature of this study was to match the ground-based
data with the IRAS observations. Temporal matching of observations is a critical
step in minimizing errors because these stars are Mira variables which have
K-band magnitude variations of up to one magnitude. The multi-epoch data allowed
for a period-luminosity relation to be determined for these Miras which was
used to calculate distances to these Galactic stars. The mass loss rates were
determined from modeling based on a modified Reimers' Formula \citep{Reimers}.
We used the 12\( \mu  \)m IRAS flux for each of the mass-losing AGB stars and
subtracted 38\% of the period-mean K-band flux, corresponding to the blackbody
contribution at 12\( \mu  \)m.  A clear linear relationship of this 12\( \mu  \)m excess,
scaled with the square of the distance can be seen
in Figure \ref{Stars Mdot} which is fitted by 
\begin{equation}
log \dot{M} = log(E_{12}*d^2) - (5.04\pm 0.02).
\end{equation}
This relationship is extrapolated to our galaxies and presented in Table \ref{K-[12]},
in the column labeled ``W94''.

\subsubsection{\label{Mdot Jacco}LMC AGB Stars}

To avoid the problem of distance uncertainties, \citet{Trams  Obs} observed
AGB stars in the LMC with a similar ground-based monitoring program to that
used by W94. The space-based IR data were matched in time with
the ground-based observations to minimize pulsation phase errors during analysis. 
Most of the observations were taken with ISOCAM in the lw10 band, with a smaller
number of observations taken with ISOPHOT. About half of the observations had
ISOCAM CVF spectra to aid in the mass loss determinations by modeling the silicate
9.7\( \mu  \)m feature. Otherwise, mass loss rates were determined solely from
SED modeling \citep{Jacco Mdot}. We used the data provided to derive the 12\( \mu  \)m
excess, which is correlated with the mass loss, leading to a linear relation
of the form 
\begin{equation}
log \dot{M} = log(E_{12}*d^2) - (4.62\pm 0.08).
\end{equation}
This relationship (Figure \ref{Stars Mdot}) provides an additional mass loss measurement for
our data (Table \ref{K-[12]}, column labeled ``L99'').

\subsubsection{\label{comparemdot}Comparison of Mass Loss Determination in the Sample Galaxies}

The three different mass loss relationships yield moderately different rates,
with a range of 0.6 dex in the zero point between the W94 relation
to the KGW relation, with L99 falling in between. We
believe that this range represents the current uncertainties in AGB mass loss
models, and, most likely, the poorly determined factors such as metallicity
and age are creating some of the dispersion. It is thought that younger AGB
populations will evolve off the main sequence faster and lead to higher mass
loss rates. On the other hand, the more metal rich an AGB star is, the more
easily it will form dust grains; for a given amount of 12\( \mu  \)m excess,
the derived mass loss will be higher in a metal rich AGB star. Both the KGW
and W94 studies sample high galactic latitude stars
 ($|b|\geq( 30^{\circ }$),
which should be older, more metal rich stars than  young
and metal poor stars in the LMC (the L99). Both factors imply that the LMC AGB stars should
have a higher mass loss rate for a given 12\( \mu  \)m excess, but the derived
rate falls between the two studies. Clearly, there is a factor of a few uncertainty
in the community about mass loss determinations. However, for similar populations
such as our early-type galaxies, this problem applies only to absolute mass
loss and the relative rates between our galaxies should be accurate to \( \sim  \)30\%
as the sum total of the error propagation reports in Table \ref{K-[12]}.

If there is a universal mass loss rate that varies linearly with the number of stars in a population, 
then galaxies' observed mass loss rates should scale with luminosity to a common
value.  When we propogate the observed errors from the various mass loss determinations and scale
by the luminosity of the galaxies (Table \ref{Galaxy Properties Table}), 
the 
range of values over the nine galaxies varies by a factor of ten.  Meanwhile, the propogated errors of
observational uncertainties predict a scatter of approximately one-half.  We could justifiably
throw out NGC 1344 and/or NGC 5102 from our sample, however these removals do not change the results.
Because the range of luminosity scaled mass loss rates is much larger than the observational uncertainties,
the data suggests that we are detecting physical differences between these populations and not 
observing the same phenomena over different sized populations.  The most likely physical differences are
age and metallicity.

\subsection{\label{theory}Theoretical Mass Loss Rates}

It is also possible to address the global mass loss rate from
theoretical stellar evolutionary models for a stellar population.
Because elliptical galaxies consist of an old coeval
population and all these stars have similar end states, it is possible to construct
arguments and/or models predicting the result from a collection of these stars
moving through their AGB phase. The first task in such a theoretical line of
reasoning is to  determine the fractional number of stars in a given old stellar
population that is currently in the AGB phase. One method for determining this
critical input parameter is to use observations of supernovae (SNe) and planetary
nebulae (PN). These stellar evolutionary phases are short lived, but over a
large enough population such as a galaxy, there are enough of these seen at
any given time to give an idea of the relative numbers of stars in these post-AGB
phases. If the number of these observables are extrapolated to include the entire
AGB phase, then an estimate can be obtained of the fraction of stars of a population
that are currently losing mass. From stellar modeling and stellar remnant observations,
it is known that stars with masses around \( 1M_{\sun } \) lose approximately
\( 0.3M_{\sun } \) in the AGB phase. Also, from stellar modeling applied to
observations of clusters of stars it is known that the approximate time spent
in the AGB phase is on the order of \( 10^{5} \) years. Combining the fraction
of a population in the AGB phase with the total mass loss for 1.0 \( M_{\sun } \)
stars and the time taken to lose this mass, it is possible to derive estimates
of galaxy-wide mass loss rates. 

\citet{Faber & Gallagher} made such arguments based on the
observations at the time and derived a scaling law of 0.015\( (M_{\sun }/10^{9}L_{\sun ,pg}) \),
where \( pg \) represents the ``photographic'' band of the international
photographic system \citep{Ref N}. Converting \( L_{pg} \) to the UBV photometric
system for K and M-type stars drops this constant by a factor of \( \sim  \)2.5-3.0
\citep{Ref O,Ref P}. With refined stellar models, it is possible to further
tune this relation by examining the fraction of stars theoretically expected
to be in an AGB phase at a given age. A few such studies have been carried out and
have slightly increased this theoretical mass loss rate by about 30\% for a
15Gyr population \citep{Theo 1,Theo 2,Theo 3,Theo 4,Theo 5,Mdot theory review}.
These adjustments lead to a relation of the form: 0.0078\( (M_{\sun }/10^{9}L_{\sun ,B}) \),
which we apply to our galaxies and report in Table \ref{K-[12]} under the column
``theory''. The theoretical rates are similar to the average of the three
different methods used to derive the observed mass loss rates. We are encouraged
by the fact that the two galaxies with the lowest predicted mass loss rates,
NGC 1344 and NGC 5102, are the two galaxies observed to have the least amount
of 12\( \mu  \)m excess emission. The small sample of galaxies precludes us
from drawing conclusions about any discrepancies between the observed and theoretical
rates, but it is natural to believe that differences will arise from metallicity
and age changes between the galaxies as they do in individual AGB stars. Rather
than viewing this as a hindrance to the study of mass loss, the relationships,
once determined, can be inverted and used as observational tools to determine
age and metallicity.

\subsection{\label{Light Profile Section}Light Profiles}

\begin{figure}
\plotone{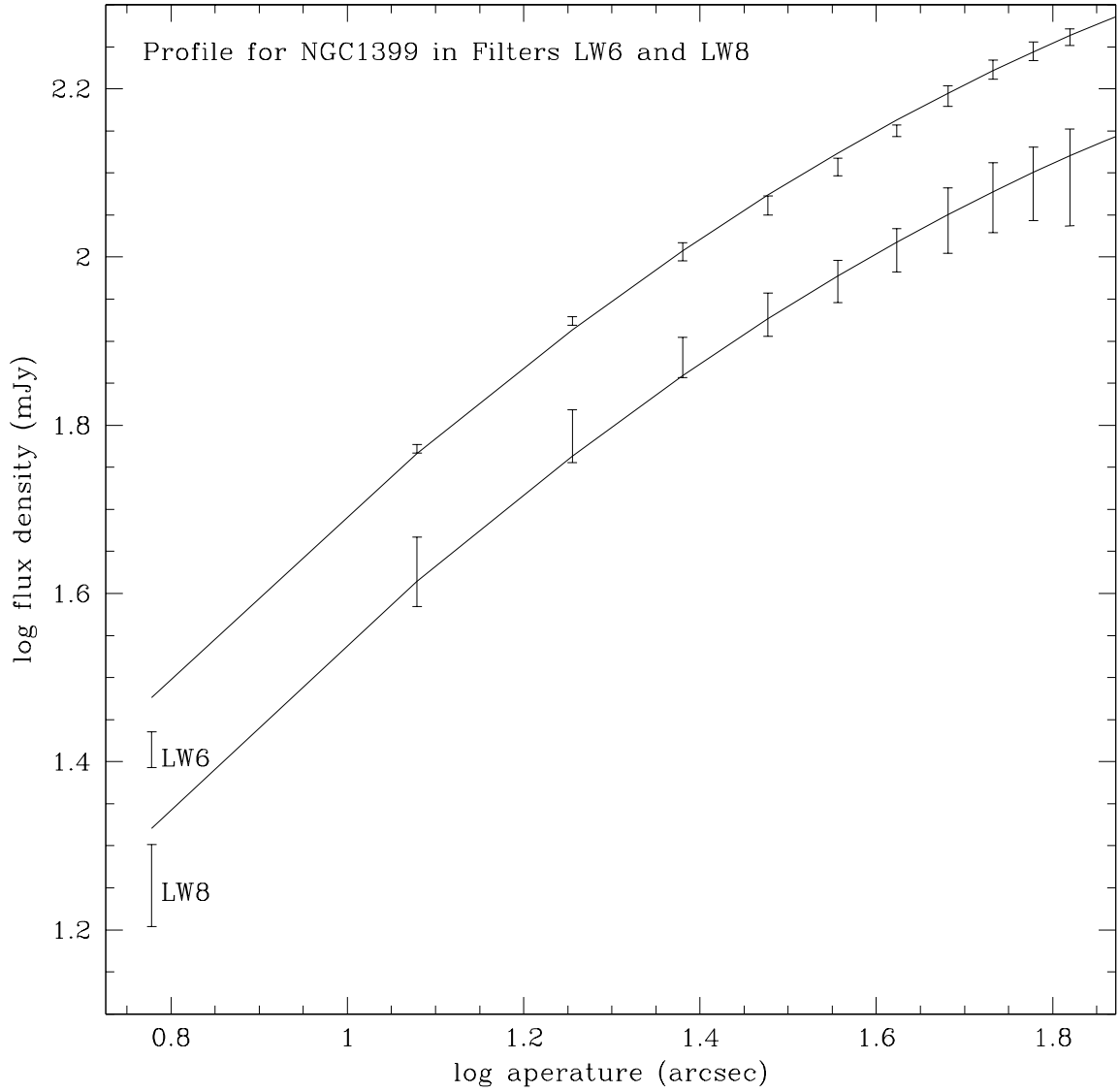}
\caption{\label{Profile+Stars}Surface brightness profiles plotted with 
de Vaucouleurs' R\protect\( ^{1/4}\protect \) law fits. K and M starlight is traced in the 
LW6 band, 7.00-8.50\protect\( \mu \protect \)m. Dust in addition to starlight is traced in the LW8 band, 10.7-12.0\protect\( \mu \protect \)m.
The agreement between the model and the data in both the LW6 and LW8 band shows that the dust 
is distributed throughout the galaxy in the same manner as the stars.
}
\end{figure}

The surface brightness profiles of our data in
all wavelength bands agree with a de Vaucouleurs'
\( R^{1/4} \) profile (See Figures \ref{Profile+Stars})
\citep{de Vaucouleurs}. A de Vaucouleurs' \( R^{1/4} \) profile fits both
the stellar bands (lw4 - lw6) as well as the stellar + dust bands (lw7 -lw9).
For our brightest galaxy, NGC 1399, where the errors are smallest and best constrain
the solution, 
the data marginally match with the model, with reduced 
\( \chi ^{2} \) of \( \sim  \)3 for 10 degrees of freedom. 
When we remove the
first data point from the fit, where there are pixelization effects
from tracking a steep profile with large pixels, 
we obtain
a reduced \( \chi ^{2} \) value of \( \sim  \)1.8, indicating a high 
correlation
between the data and the model. The model remains consistent with the data when
we fix \( r_{e} \), the effective radius, to the optically determined value
(reported in Table \ref{Galaxy Properties Table}) rather than have it vary
as a free parameter. In the redder dust bands, the flux profile is again consistent
with a \( R^{1/4} \) profile, with reduced \( \chi ^{2} \) at or below unity.
The stellar bands (lw4-lw6) are expected to fit a \( R^{1/4} \) profile as
K and M stars are the only source of radiation in this region. This provides
a check on the data, confirming that the data processing removed all large scale
variations over the field of view. Because the stellar + dust bands (lw7-lw9)
are also consistent with a \( R^{1/4} \)profile, we conclude that the MIR excess
originates from the same population as the K and M stars. More specifically,
the MIR excess originates from the dust in the circumstellar shells of the subset
of the K and M stars that are currently going through their AGB phase.

\section{\label{conclusion}Summary and Concluding Remarks}

Nine elliptical galaxies were observed with CAM on ISO in six narrow bands between 6\( \mu  \)m 
and 15\( \mu  \)m.
From 6\( \mu  \)m to 9\( \mu  \)m the emission is consistent with the combined stellar emission
from the K and M-type stars that dominate elliptical galaxy's integrated light.
In eight of these galaxies we 
detected excess over stellar photospheric emission from 9\( \mu  \)m to 15\( \mu  \)m.
For one galaxy, NGC 1404, ISOCAM CVF data, with its finer 
spectral resolution, shows the excess emission is 
consistent with the known 9.7\( \mu  \)m oxygen-rich AGB silicate feature. 
We used Galactic and LMC AGB stars to calibrate a scaling relation, revealing 
galactic-wide mass loss rates for these galaxies. These observed rates mostly agree with 
theoretical predictions.  The observed rates do not scale with luminosity to a universal rate 
and thus suggests physical differences between these populations.
We also show that emission at all wavelengths is consistent with a de Vaucouleurs' \( R^{1/4} \)
law.

Now that it is possible
to observe the signatures of mass loss, it would be valuable
to revisit the predicted mass loss characteristics of cluster and galaxy sized populations.
Tracking the population's total mass loss through time in a quantitative manner would be 
a valuable contribution to the field.
Also the
calibration of the relationship between mass loss and mid IR excess needs further
investigation with additional studies of individual AGB stars.
In particular, the effects of different metallicites 
should be investigated, as it is known to play a role, but the exact nature is not yet 
determined. 
An exciting avenue of research that will surely develop once the calibrations of these AGB features 
is accurately known, is the Visual-MIR color-color diagnostics.  This observational tool was recently explored in a
theoretical study  
by \citet{Bressan98}, which shows that for certain populations these types of relations could potentially 
break the age-metallicity degeneracy that has long plagued optical color-color relations.  
With SOFIA, SIRTIF and other upcoming IR missions, in addition to the continuing work
though high altitude ground-based windows, it should be possible to continue to investigate
these issues.

\acknowledgements{}

We thank the ISO discretionary time committee for generously granting time for
NGC 1404 with the CVF as follow-up to our CAM data.  We would also like to thank
the IPAC support team, especially Ken Ganga who guided us through the initial reductions of the 
tricky CAM data.  This research was supported in part by the NASA/ISO Guest Observer Program and grants 
from NASA's LTSA Program.  This research has made use of the NASA/IPAC Extragalactic Database (NED) which
is operated by the Jet Propulsion Laboratory, California Institute of Technology,
under contract with the National Aeronautics and Space Administration.

\newpage

\begin{deluxetable}{cccccccc}
\tabletypesize{\scriptsize}
\tablecaption{Galaxy Properties \label{Galaxy Properties Table} }
\tablewidth{0pt}

\tablehead{
\colhead{Name} &
\colhead{RA\tablenotemark{a}} &
\colhead{Dec\tablenotemark{a}} &
\colhead{$B^{0}_{T}$\tablenotemark{b,c}} &
\colhead{D\tablenotemark{b,d}} &
\colhead{$r_{e}$\tablenotemark{b,e}} &
\colhead{$logL_{B}$\tablenotemark{f}} &
\colhead{Hubble Type\tablenotemark{b}}
\\
\colhead{} &
\multicolumn{2}{c}{(J2000)} &
\colhead{(mag)} &
\colhead{(km/s)} &
\colhead{(\arcsec)} &
\colhead{(ergs/s)}  &
\colhead{}
} 
\startdata
NGC 1344&
03 28 19.3&
-31 04 04&
11.11&
1422\( \pm  \)88&
38.38&
43.35\( \pm  \)0.06&
E5
\\
NGC 1395&
03 38 29.7&
-23 01 40&
10.94&
1990\( \pm  \)187&
45.07&
43.71\( \pm  \)0.06&
E2
\\
NGC 1399&
03 38 29.3&
-35 27 01&
10.55&
1422\( \pm  \)88&
42.37&
43.58\( \pm  \)0.06&
E1P
\\
NGC 1404&
03 38 52.0&
-35 35 34&
10.89&
1422\( \pm  \)88&
26.72&
43.44\( \pm  \)0.06&
E1
\\
NGC 1407&
03 40 11.8&
-18 34 48&
10.57&
1990\( \pm  \)187&
71.96&
43.86\( \pm  \)0.12&
E0
\\
NGC 4636&
12 42 50.0&
+02 41 17&
10.20&
1333\( \pm  \)71&
101.14&
43.66\( \pm  \)0.06&
E0
\\
NGC 5102&
13 21 57.6&
-36 37 49&
10.57&
202\( \pm  \)25&
23.29&
41.64\( \pm  \)0.12&
S0
\\
NGC 5846&
15 06 29.2&
+01 36 21&
10.67&
2336\( \pm  \)284&
82.61&
43.96\( \pm  \)0.12&
E0
\\
NGC 7507&
23 12 07.5&
-28 32 22&
11.15&
1750\( \pm  \)371&
31.41&
43.52\( \pm  \)0.06&
E0
\\
\enddata
\tablenotetext{a}{Values taken from NED (New Extragalactic Database).}
\tablenotetext{b}{\citet{Faber et al}}
\tablenotetext{c}{Total extrapolated B magnitudes assuming an $R^{1/4}$ law.}
\tablenotetext{d}{Distances corrected for solar motion and Virgo infall.}
\tablenotetext{e}{Effective radii from B-band images.}
\tablenotetext{f}{Luminosity in the B-band derived from $B^0_T$ and distances.}


\end{deluxetable}


\begin{deluxetable}{cccccccc}
\tabletypesize{\scriptsize}
\tablecaption{ISOCAM Filters and Galaxy Exposure Times\label{Exposure Time Table}}
\tablewidth{0pt}

\tablehead{
\colhead{} &
\colhead{} &

\multicolumn{6}{c}{ISOCAM LW Filters} 

\\

\colhead{} &
\colhead{} &

\colhead{LW4} &
\colhead{LW5} &
\colhead{LW6} &
\colhead{LW7} &
\colhead{LW8} &
\colhead{LW9} 
} 
\startdata
\multicolumn{2}{c}{\( \lambda _{o} \) (\( \mu  \)m) }&
6.00&
6.75&
7.75&
9.62&
11.4&
15.0\\
 
\multicolumn{2}{c}{\( \lambda _{min}-\lambda _{max} \) (\( \mu  \)m) }&
5.50-6.50&
6.50-7.00&
7.00-8.50&
8.50-10.7&
10.7-12.0&
14.0-16.0\\
\tableline
\colhead{Galaxy Name} &
\colhead{ISO Orbit} &
\multicolumn{6}{c}{Exposure Time (s) }\\
\tableline

NGC 1344&
802&
560&
350&
200&
190&
200&
180\\
 
NGC 1395&
848&
560&
360&
200&
190&
190&
170\\
 
NGC 1399&
802&
560&
350&
190&
180&
180&
170\\
 
NGC 1404&
593&
560&
360&
210&
190&
200&
180\\
 
NGC 1407&
820&
560&
360&
200&
190&
200&
170\\
 
NGC 4636&
574&
550&
360&
190&
190&
190&
170\\
 
NGC 5102&
632&
560&
350&
190&
200&
200&
170\\
 
NGC 5846&
627&
560&
360&
210&
190&
190&
180\\
 
NGC 7507&
537&
570&
360&
210&
200&
200&
170\\
\enddata 

\tablecomments{ 
Exposure times for LW4 filter include pre-observation, detector stabilization time.
}
\end{deluxetable}

\begin{deluxetable}{ccccccc}
\tabletypesize{\scriptsize}
\tablewidth{0pt}
\tablecaption{\label{fluxes}Galaxy ISOCAM 6.0-15.0\( \mu  \)m Fluxes}

\tablehead{
\colhead{Galaxy} &
\colhead{6\( \mu  \)m}&
\colhead{6.75\( \mu  \)m}&
\colhead{7.75\( \mu  \)m}&
\colhead{9.62\( \mu  \)m}&
\colhead{11.4\( \mu  \)m}&
\colhead{15.0\( \mu  \)m}
\\
\colhead{}&
\multicolumn{6}{c}{(mJy)}
}
\startdata
NGC 1344&
80.4\( \pm  \)3.2&
64.9\( \pm  \)2.1&
47.0\( \pm  \)1.9&
31.9\( \pm  \)3.5&
18.9\( \pm  \)3.9&
7.7\( \pm  \)3.6\\
NGC 1395&
121.2\( \pm  \)2.6&
97.9\( \pm  \)3.0&
74.4\( \pm  \)2.6&
60.8\( \pm  \)4.5&
42.7\( \pm  \)4.9&
26.7\( \pm  \)4.0\\
NGC 1399&
167.5\( \pm  \)3.4&
133.3\( \pm  \)5.5&
100.5\( \pm  \)1.9&
92.9\( \pm  \)4.8&
73.2\( \pm  \)3.9&
47.3\( \pm  \)3.2\\
NGC 1404&
158.7\( \pm  \)3.5&
126.8\( \pm  \)9.2&
97.0\( \pm  \)2.2&
77.5\( \pm  \)4.9&
55.9\( \pm  \)4.8&
30.8\( \pm  \)2.2\\
NGC 1407&
169.1\( \pm  \)2.8&
133.1\( \pm  \)5.6&
100.6\( \pm  \)1.9&
85.7\( \pm  \)3.2&
69.5\( \pm  \)3.9&
46.0\( \pm  \)3.2\\
NGC 4636&
114.2\( \pm  \)2.5&
85.3\( \pm  \)2.8&
73.4\( \pm  \)5.5&
55.4\( \pm  \)3.8&
59.4\( \pm  \)6.5&
40.7\( \pm  \)6.6\\
NGC 5102&
82.9\( \pm  \)1.6&
78.9\( \pm  \)3.9&
53.7\( \pm  \)4.9&
36.0\( \pm  \)4.3&
35.8\( \pm  \)6.5&
17.5\( \pm  \)5.2\\
NGC 5846&
97.8\( \pm  \)2.8&
90.4\( \pm  \)10.3&
60.4\( \pm  \)2.7&
52.0\( \pm  \)4.5&
33.6\( \pm  \)6.5&
21.1\( \pm  \)5.1\\
NGC 7507&
109.7\( \pm  \)1.9&
85.6\( \pm  \)2.9&
63.8\( \pm  \)6.2&
72.1\( \pm  \)5.7&
41.4\( \pm  \)6.4&
32.2\( \pm  \)5.9\\
\enddata

\tablecomments{ 
Fluxes are the mean of four raster aperature photometry measurements taken at 30 arcseconds radius centered on 
the galaxy with background subtraction 
from the outer regions of the image. 
Errors are the larger of the standard deviation of the mean of the four raster measurments or the statistical
noise in the sky region of image.
}


\end{deluxetable}


%

\begin{deluxetable}{cccc}

\tabletypesize{\scriptsize}
\tablecaption{NGC 1404 ISOCAM CVF Fluxes\label{fluxes-cvf}}
\tablewidth{0pt}
\tablehead{
\colhead{Wavelength(\( \mu  \)m)} &
\colhead{Flux(mJy)} &
\colhead{Wavelength(\( \mu  \)m)} &
\colhead{Flux(mJy)} }
\startdata
6.044&
84.98\( \pm  \)2.90&
8.993&
52.09\( \pm  \)3.38\\
6.221&
80.57\( \pm  \)2.94&
9.291&
49.26\( \pm  \)3.77\\
6.396&
85.49\( \pm  \)2.78&
9.660&
50.64\( \pm  \)4.44\\
6.569&
80.29\( \pm  \)2.96&
9.986&
50.78\( \pm  \)3.10\\
6.741&
72.94\( \pm  \)2.85&
10.31&
47.27\( \pm  \)3.56\\
6.911&
70.24\( \pm  \)3.14&
10.63&
42.06\( \pm  \)3.96\\
7.080&
66.57\( \pm  \)3.74&
10.95&
38.08\( \pm  \)4.83\\
7.248&
69.21\( \pm  \)2.59&
11.27&
37.41\( \pm  \)4.39\\
7.414&
63.61\( \pm  \)3.80&
11.58&
30.16\( \pm  \)4.30\\
7.578&
62.84\( \pm  \)3.35&
11.89&
27.54\( \pm  \)4.24\\
7.741&
55.83\( \pm  \)2.88&
12.21&
21.11\( \pm  \)4.61\\
7.903&
54.72\( \pm  \)2.91&
12.51&
20.76\( \pm  \)3.88\\
8.063&
62.78\( \pm  \)3.25&
12.82&
31.47\( \pm  \)4.49\\
8.222&
56.73\( \pm  \)3.13&
13.12&
32.15\( \pm  \)4.32\\
8.379&
50.94\( \pm  \)2.45&
13.43&
26.96\( \pm  \)5.04\\
8.534&
52.31\( \pm  \)2.97&
13.73&
26.07\( \pm  \)4.97\\
8.689&
46.61\( \pm  \)2.84&
14.02&
29.46\( \pm  \)4.50\\
8.842&
46.93\( \pm  \)3.22&
&
\\

\enddata
\tablecomments{ 
Fluxes are the taken at 30 arcseconds radius centered on 
the galaxy with background subtraction 
from the outer regions of the image. 
Errors are the statistical
noise in the sky region of image.
}

\end{deluxetable}


\begin{deluxetable}{ccccccccc}
\tabletypesize{\scriptsize}
\tablecaption{Galaxy Excess MIR Emission and Mass Loss Rates\label{K-[12]}}
\tablewidth{0pt}

\tablehead{
\colhead{Galaxy}&
\colhead{K-[12]\tablenotemark{a}}&
\colhead{$L^{30\arcsec}_{B}/L_{B}$\tablenotemark{b}}&
\colhead{$E_{12}$\tablenotemark{c}}&
\colhead{$log (E_{12}*d^2)$\tablenotemark{d}}&
\colhead{$\dot{M}$ KGW}&
\colhead{$\dot{M}$ W94}&
\colhead{$\dot{M}$ L99}&
\colhead{$\dot{M}_{theory}$} 
\\
\colhead{} &
\colhead{} &
\colhead{} &
\colhead{$(mJy)$}&
\colhead{$(log (mJy*Mpc^2))$}&
\colhead{$(M_{\sun}/yr)$}&
\colhead{$(M_{\sun}/yr)$}&
\colhead{$(M_{\sun}/yr)$}&
\colhead{$(M_{\sun}/yr)$}
}

\startdata
NGC 1344&
-0.09\( \pm  \)0.01&
0.4327&
-4.5\( \pm  \)1.0&
\( <  \)3.45\( \pm  \)0.07&
\( <  \)0.10\( \pm  \)0.04&
\( <  \)0.03\( \pm  \)0.01&
\( <   \)0.07\( \pm  \)0.02&
0.37\\

NGC 1395&
0.26\( \pm  \)0.01&
0.3905&
25.2\( \pm  \)2.6&
4.37\( \pm  \)0.06&
0.86\( \pm  \)0.40&
0.21\( \pm  \)0.06&
0.57\( \pm  \)0.22&
0.86\\
 
NGC 1399&
0.45\( \pm  \)0.01&
0.4065&
63.0\( \pm  \)3.2&
4.48\( \pm  \)0.06&
1.10\( \pm  \)0.40&
0.27\( \pm  \)0.05&
0.73\( \pm  \)0.22&
0.64\\
 
NGC 1404&
0.25\( \pm  \)0.01&
0.5322&
22.4\( \pm  \)1.7&
4.03\( \pm  \)0.06&
0.39\( \pm  \)0.15&
0.10\( \pm  \)0.02&
0.26\( \pm  \)0.08&
0.46\\
 
NGC 1407&
0.40\( \pm  \)0.01&
0.2787&
77.7\( \pm  \)4.0&
4.86\( \pm  \)0.08&
2.64\( \pm  \)1.13&
0.66\( \pm  \)0.16&
1.75\( \pm  \)0.65&
1.21\\
 
NGC 4636&
0.56\( \pm  \)0.03&
0.2107&
107.3\( \pm  \)10.6&
4.65\( \pm  \)0.06&
1.64\( \pm  \)0.62&
0.41\( \pm  \)0.08&
1.09\( \pm  \)0.35&
0.76\\
 
NGC 5102&
0.26\( \pm  \)0.02&
0.5705&
12.6\( \pm  \)2.1&
2.09\( \pm  \)0.13&
0.004\( \pm  \)0.002&
0.001\( \pm  \)0.001&
0.003\( \pm  \)0.001&
0.007\\
 
NGC 5846&
0.21\( \pm  \)0.01&
0.2497&
26.2\( \pm  \)4.4&
4.53\( \pm  \)0.13&
1.22\( \pm  \)0.65&
0.31\( \pm  \)0.11&
0.82\( \pm  \)0.38&
1.52\\
 
NGC 7507&
0.45\( \pm  \)0.03&
0.4891&
33.3\( \pm  \)4.4&
4.38\( \pm  \)0.19&
0.88\( \pm  \)0.59&
0.22\( \pm  \)0.11&
0.58\( \pm  \)0.36&
0.55\\
\enddata 

\tablenotetext{a}{K-band relative flux calculated assuming 6.00-7.75\( \mu  \)m is stellar emission and 
employing the Rayleigh-Jeans law.  12\( \mu  \)m magnitude, [12], derived from linear combination of narrow band filters.
Zero point for K-[12] relation set by hot blackbody (T$\sim$10,000K). }
\tablenotetext{b}{Fraction of flux encircled at 30\( \arcsec  \) assuming effective radii listed in Table \ref{Galaxy Properties Table} and a de Vaucouleurs profile. }
\tablenotetext{c}{Expected excess emission over a blackbody at 12\( \mu  \)m for the entire galaxy, assuming emission 
follows a de Vaucouleurs profile. (See Section \ref{Light Profile Section}). }
\tablenotetext{d}{The 12\( \mu  \)m excess is scaled by distance squared (with $H_o=65 km s^{-1}Mpc^{-1}$)
and reported in the log. 
 }

\end{deluxetable}

\newpage

\clearpage

\begin{figure}
\plotone{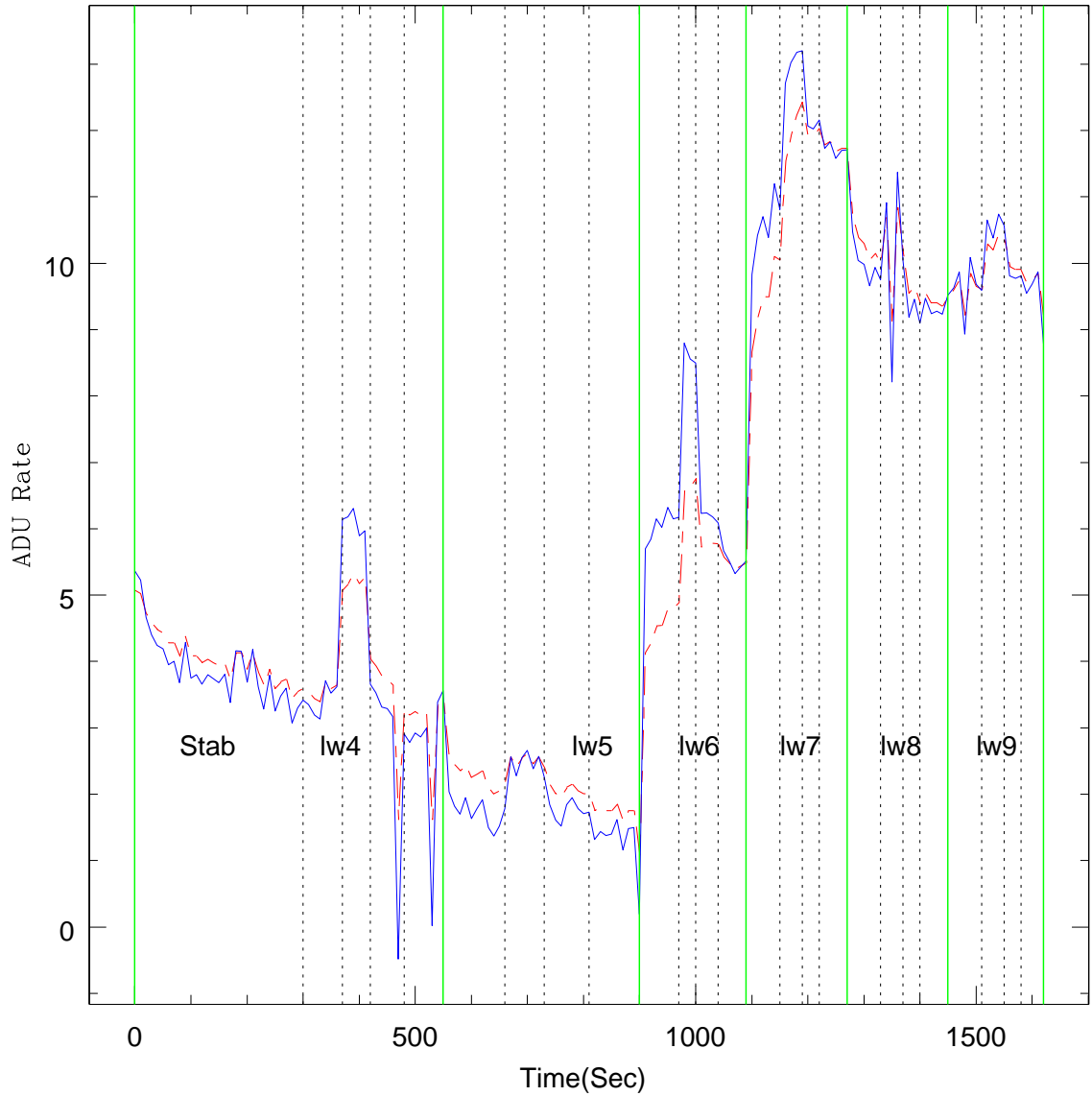} 
\caption{\label{source-trans}The electronic detection history of a typical source pixel.
Same graphical representation as Figure \ref{sky-trans}.
With the source pixel electronic history it can be
seen that the timescale for flux changes is even smaller as the satellite dithers
about the center of the galaxies in different pointings.
}

\end{figure}

\clearpage
\begin{figure}
\plotone{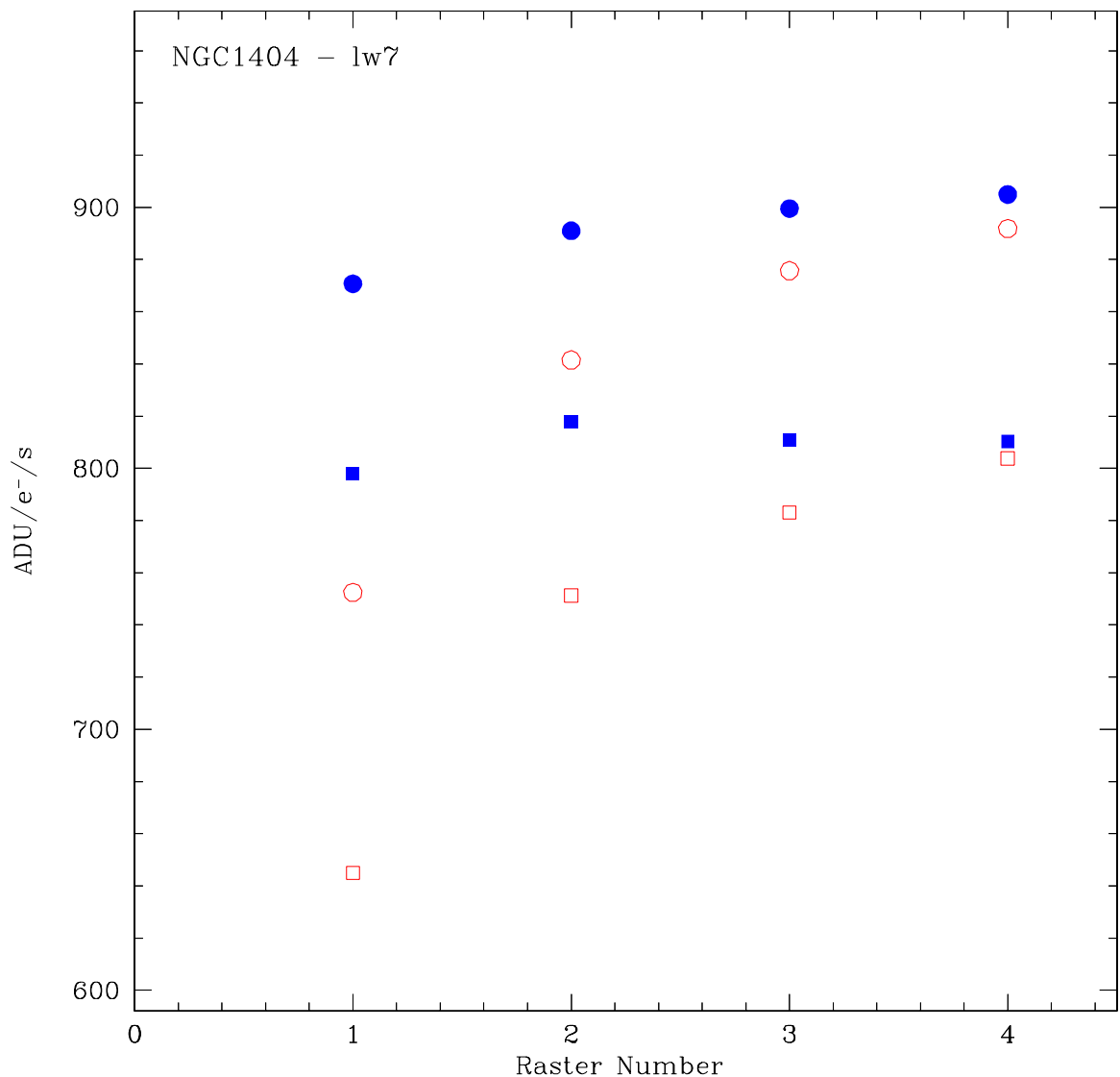} 
\caption{\label{group-trans}Transient corrected (filled points) and raw (open points)
data in source (circles) and background (squares) regions for NGC1404 in filter
lw7. It appears the detector has reached a stablized value by the end of the fouth raster observation, while
it remains transient in the first two pointings.
The flux of the second pointing is 15\% larger than the
first, in both source and sky. While the third and fourth pointings have
a 3\% difference. Also the standard deviation of the mean is much larger in
the un-corrected data, with a clear positive derivative present in the data.  The corrected data for all pointings 
is near the stablized values of the un-corrected third and fourth pointings.
}

\end{figure}

\clearpage

\clearpage

\clearpage

\clearpage

\end{document}